\documentclass[aps,prb,preprint,superscriptaddress, showpacs]{revtex4-1}
\usepackage{placeins}
\usepackage{amsmath}
\usepackage{amssymb}
\usepackage{amsthm}
\usepackage[dvips]{graphicx}
\usepackage{pslatex}
\usepackage{gensymb}

\begin{document}
\bibliographystyle{apsrev4-1}
%
\title{Anomalously Weak Cooper Pair-breaking by Exchange Energy in Ferromagnet/Superconductor Bilayers}
%
%
\author{T. R. Lemberger}
\email{trl@physics.osu.edu}
\affiliation{Dept. of Physics, Ohio State Univ., Columbus, Ohio 43210, USA}
\author{M. J. Hinton} 
\email{hinton.84@osu.edu}
\affiliation{Dept. of Physics, Ohio State Univ., Columbus, Ohio 43210, USA}
\author{Jie Yong}
\affiliation{Dept. of Physics, Ohio State Univ., Columbus, Ohio 43210, USA}
\author{J. M. Lucy} 
\affiliation{Dept. of Physics, Ohio State Univ., Columbus, Ohio 43210, USA}
\author{A. J. Hauser} 
\affiliation{Dept. of Physics, Ohio State Univ., Columbus, Ohio 43210, USA}
\author{F. Y. Yang} 
\affiliation{Dept. of Physics, Ohio State Univ., Columbus, Ohio 43210, USA}
%
%
%
%
%
\date{\today}
\begin{abstract}
We report the superconducting transition temperature $T_c$ vs. thickness $d_F$ of Ferromagnet/Superconductor (F/S) bilayers, where F is a strong $3d$ ferromagnet (Ni, Ni$_{0.81}$Fe$_{0.19}$ (Permalloy),  Co$_{0.5}$Fe$_{0.5}$) and S = Nb, taken from superfluid density measurements rather than resistivity. By regrouping the many physical parameters that appear in theory, we show that the effective exchange energy is determined from the F film thickness $d_F$ where $T_c$ vs. $d_F$ begins to flatten out. Using this rearranged theory we conclude: 1) the effective exchange energy, $E_{ex}$, is about 15 times smaller than measured by ARPES and 5 times smaller than deduced in previous studies similar to ours; 2) the dirty-limit coherence length, $\xi_{F}$, for Cooper pairs in F is larger than the electron mean free path, $\ell_F$; and 3) the $3d$-F/Nb interface is enough of a barrier that Cooper pairs typically must hit it several times before getting through. The Py/Nb and CoFe/Nb interfaces are more transparent than the Ni/Nb interface.
\end{abstract}
%
%
\maketitle
\section{INTRODUCTION}

There is a great deal of interesting physics involved with Cooper pairs that are introduced into ferromagnets (F) through the proximity effect.[e.g., \cite{Buzdin2005}]  Hallmarks of this physics include an overall exponential decay of the order parameter into the ferromagnet on a much shorter length scale than in nonmagnetic metals, oscillations in the order parameter that accompany the exponential decay, and possible generation of spin-triplet Cooper pairs, which are not as strongly affected by the ferromagnetic exchange energy as s-wave pairs are. Naturally, much work has focused on the remarkable qualitative features of order parameter oscillations and spin-triplet pairs. The present work focuses on the quantitative issue of the effective exchange energy, $E_{ex}$, experienced by s-wave Cooper pairs in strong 3-d ferromagnets.

We take a critical look at effective exchange energies that are deduced from the simplest experiments, namely, measurements of $T_c$ vs. $d_F$ in S/F bilayers involving $3d$ ferromagnets. We present our own results on Nb/Ni, Nb/CoFe, and Nb/Py (Permalloy) bilayers, where $T_c$ is obtained from superfluid density measurements rather than resistivity, which is the standard method. Previous similar studies have found that $E_{ex}$ is about equal to $k_BT_C$, where $T_C$ is the Curie temperature. For example, Sidorenko et al.\cite{Sidorenko2003} studied Nb/Ni bilayers, Aarts et al.\cite{Aarts1997} studied $V$/$V_{1-x}Fe_x$ bilayers, and Kim et al.\cite{Kim2005} studied Nb/Ni and Nb/CoFe bilayers. The latter already emphasized that $k_BT_C$ is several times smaller than measured and calculated values of exchange energy, so it is not so surprising that we find the effective exchange energy felt by Cooper pairs in F to be smaller by another order of magnitude. 

Available theory ignores the fact that F has strong electron correlations, and that densities of states and Fermi velocities of spin-up and spin-down electrons generally differ. Thus, as emphasized by Tagirov,\cite{Tagirov1998} the quantitative accuracy of theory is suspect. However, maybe it is accurate, if we can figure out what density of states and Fermi velocity to use for F. There are experimental results that guide us, e.g., ARPES yields Fermi velocities in pure Ni and in alloy Py, and finds no difference between spin-up and spin-down bands.\cite{Petrovykh1998}  

We analyze $T_c$ vs. $d_F$ taken from our superfluid density measurements on: Ni/Nb, Py/Nb [Py = Ni$_{0.81}$Fe$_{0.19}$], and CoFe/Nb [Co$_{0.5}$Fe$_{0.5}$] bilayers, and we analyze published data on Ni/Nb and CoFe/Nb bilayers for comparison. Since our goal is to extract an effective exchange energy, at the outset we must identify values for other important parameters. The electron density of states, Fermi velocity, and bulk resistivity of Nb are reliably found in the literature. For $3d$ ferromagnets, the literature on these properties is less extensive than for Nb, but it is sufficient to place a reliable lower limit on the effective density of states at the Fermi level. Also, it provides Fermi velocities. The single most important materials parameter is the ratio of densities of states, $2N_F(0)/2N_{Nb}(0)$, because it alone is needed to determine $E_{ex}$ from the data. As a check on the overall reasonableness of the fit parameters, we expect to find that Cooper pairs in F must hit the F/S interface at least a few times in order to get through. After all, one ``hit'' is required even if the interface is perfectly transparent. Differences in Fermi velocities plus a bit of disorder and interdiffusion at the interface should bump up the required number of hits to at least a few.

\section{THEORY}

In this section we examine existing theoretical results with two purposes. First, we want to show that the basic physics involved in $T_c$ vs. $d_F$ for S/F bilayers can be understood pretty clearly by imagining that the s-wave cooper pairs in F are formed of time-reversed electronic states, (opposite momenta and spins), just like they are in S, so that electrons in a pair have energies that differ by the effective exchange energy. That this viewpoint works is surprising because theory is formulated in terms of Cooper pairs of equal-energy electrons (at the Fermi surfaces of spin-up and spin-down electrons). Second, we want to make notational connections between different formulations of theory that have been employed by different experimental groups in fitting data. For example, Houzet and Meyer (HM) characterize the F/S interface with a specific resistance, $R_b$, while Tagirov uses a transmission parameter, $T_F$, that runs from 0 to infinity. We will see that $R_b$ is equivalent to $2 \rho_F \ell_F/3T_F$, where the ``rho-ell'' product is: $\rho \ell \equiv 3/2N(0)v_Fe^2$.
 
Physically, the suppression of $T_c$ in bilayers is determined by the net flux of Cooper pairs out of S. A typical pair moving around in S occasionally bumps into the S/F interface, and, after a few bumps, it transits into F where it dephases at a rate $E_{ex}/\hbar$ due to the energy difference between electrons in a Cooper pair. We emphasize that $E_{ex}$ represents the effective exchange energy felt by the pair. While in F, the pair occasionally bumps into the F/S interface. Unsurprisingly, theory finds that when F is thin, pairs simply bounce ballistically back and forth between the two surfaces of F, each round trip having a length of about $3d_F$. After a few bumps, and if the dephasing angle $\Gamma_F \equiv \Delta t E_{ex}/\hbar$ has not exceeded about $\pi$ during the time $\Delta t$ that the pair has lingered in F, then the pair returns to S weakened, perhaps, but not broken. 

It is sufficient for our purpose to compare the notation of Houzet and Meyer (HM) formulation of dirty-F theory with the more general formulation of Tagirov which covers the range from clean to dirty F. We begin with the HM version of the equation for $T_c$ as a function of the complex pair-breaking rate, $1/\tau_S$, which depends on $\Gamma_F$, and thereby on $d_F$: \cite{Houzet2009}
\begin{equation}
\ln\left(\frac{T_c}{T_{c0}}\right)= \Psi\left(\frac{1}{2}\right) - \Re\left\{ \Psi\left(\frac{
1}{2}+\frac{1}{2\pi T_c\tau_S}\right) \right\}
\label{Eq1} \\=\Re\left\{\sum_{n=0}^{\infty}\frac{-1/2\pi T_c\tau_S}{(n+1/2)(n+1/2+1/2\pi T_c \tau_S)}\right\},
\end{equation}
where $\Psi(x)$ is the digamma function. We use this equation to fit our data, but for simplicity we restrict the present discussion to F/S bilayers in which the suppression of $T_c$ is small. In this case, Eq.~\ref{Eq1} simplifies to:
\begin{equation}
T_c \approx T_{c0} - \frac{\pi\hbar}{4k_B} \Re\left\{\frac{1}{\tau_S}\right\}.
\label{Eq2}
\end{equation}
There are many ways to express $1/\tau_S$ in terms of parameters like coherence length, $\xi_S$, and resistivity, $\rho_S$. Using various free-electron relations, we choose to write is as: 

\begin{equation}
\frac{1}{\tau_S}=\frac{1}{2\tau_{tun,S}} \frac{1}{1+1/\sqrt{2i}(R_b/\rho_F \xi_F)\tanh(\sqrt{2i}d_F/\xi_F)}.
\label{Eq3}
\end{equation}
$\rho_F$ in Eq.~\ref{Eq3} is the resistivity of F. $\xi_F \equiv \sqrt{\hbar D_F/E_{ex}}$ in Eq. \ref{Eq3} is the mean-free-path-limited coherence length for s-wave Cooper pairs in F. Tagirov uses ``$\xi_F$'' to represent the intrinsic clean-F coherence length, which we denote here as, $\xi_{F0} \equiv \hbar v_{FF}/E_{ex}$, to make the distinction clear.
Because the density of states in Nb is known, we write the ``tunneling'' rate in Eq.~\ref{Eq3} as:
\begin{equation}
\frac{1}{2\tau_{tun,S}} \equiv \frac{1}{4N_S(0)e^2d_SR_b}.
\label{Eq4}
\end{equation}
$2\tau_{tun,S}$ is the time a typical Cooper pair spends in S before jumping into F, when S is thin. \cite{Fominov2002} It sets the scale for pair-breaking because it is the net pair-breaking rate if none of the pairs returns from F to S before breaking. The appellation ``tunneling'' arises because Eq.~\ref{Eq4} also describes pair-breaking in superconductor-insulator-normal metal tunnel junctions.\cite{McMillan1968, Lemberger1984, Lemberger1984a} 

There is ambiguity in the resistivity $\rho_F$ in Eq.~\ref{Eq3}. Is it the in-plane resistivity which involves surface scattering and scattering from grain boundaries and misfit dislocations, or the resistivity perpendicular to the film, which seems perhaps more relevant to diffusion of electrons into F. Rather than spend time discussing this point, we avoid it by eliminating $\rho_F$ from the theory.

At this point we pause to compare with Tagirov's theory, as simplified in Sidorenko et al. \cite{Sidorenko2003}. Tagirov's equation for $T_c$ is the same as Eq.~\ref{Eq1}, but with $2\pi T_{c0} \rho$ in place of $1/\tau_S$, where $\rho = \frac{\xi_{S0}^2}{2d_S^2}\phi^2$ and an expression is given for $\phi \tan(\phi)$. (As noted above, we are substituting ``$\xi_{F0}$'' in place of ``$\xi_F$'' in Tagirov.) When pair-breaking is weak, which is the case considered in this section, $\phi$ is small, and Tagirov's expression for $\phi \tan(\phi)$ immediately yields $\phi^2$. Using this, Tagirov's version of $1/\tau_S$ becomes:
\begin{equation}
\left.{\frac{1}{\tau_S}} \right\vert_{Tagirov}=\frac{1}{4N_{S}(0) d_S e^2 \left( 2 \rho_F \ell_F / 3T_F \right)} \frac{1}{1+T_F i k_F \xi_F / 2 \tanh{\left(k_F d_F \right)}},
\label{Tagirov}
\end{equation}
where $k_F \equiv \sqrt{i\xi_{F0}/\ell_F - 1}/\xi_{F0}$. The similarity with HM Eqs.~\ref{Eq3} and~\ref{Eq4} is obvious. We note only that $2\rho_F \ell_F/3T_F$ in Tagirov is equivalent to $R_b$ in HM. The transmission coefficient $T_F$ runs from 0 to infinity, corresponding to $R_b$ running from infinity to 0. As a practical matter, differences in Fermi velocities between S and F, plus a bit of lattice-mismatch disorder and interdiffusion at the F/S interface mean that we can expect $T_F$ to be less than, say, 1/2, so that Cooper pairs must hit the F/S interface several times before getting through. 
 
Returning to HM, ferromagnetism enters through the factor multiplying $1/2\tau_{tun,S}$ in Eq.~\ref{Eq3}. This is where the dependence of $T_c$ on $d_F$ emerges. In the thin-F regime, $d_F \ll \xi_F$, the $\tanh$ in Eq.~\ref{Eq3} equals its argument, and the following simplification results:
\begin{equation}
\sqrt{2i}\frac{R_b}{\rho_F \xi_F}\tanh{\frac{\sqrt{2i}d_F}{\xi_F}} \approx i \frac{E_{ex}}{\hbar} \frac{4d_F}{v_{FF}} \frac{3R_b}{2\rho_F \ell_F}  = i\Gamma_F,
\label{Eq5}
\end{equation}
We identify this expression as the dephasing angle, $\Gamma_F$, of a Cooper pair because it is the dephasing rate, $E_{ex}/\hbar$, times the dwell time in F, i.e., the time, $\Delta t\approx 4d_F/v_{FF}$, that a typical pair takes to bounce off the back of F and return to the F/S interface, multiplied by the typical number of hits needed to get through the F/S interface, $\approx 3R_b/2\rho_F \ell_F$. \cite{Mancusi2011, Tagirov1998, Lazar2000} From this equation, we see that $\xi_{F0}$ is the distance a Cooper pair travels in F while accumulating a dephasing angle of unity: $\Gamma_F = (\xi_{F0}/v_{FF}) \times (E_{ex}/\hbar) = 1$. 
Free-electron relations like $\rho = 1/2N(0)De^2$ ($D$ = diffusion constant) allow us to express $\Gamma_F$ in terms of only $E_{ex}$ and the density of states, $2N_F(0)$:
\begin{equation}
\Gamma_F = 2\tau_{tun,F}E_{ex}/\hbar,
\label{Eq6}
\end{equation}
where $2\tau_{tun,F}=4N_F(0)e^2d_FR_b$ has the same form as $2\tau_{tun,S}$, but with $2N_F(0)d_F$ in place of $2N_S(0)d_S$. Thus, $2\tau_{tun,F}$ is the time a typical Cooper pair spends in F before jumping back into S, when F is thin.

Putting this all together, Eq.~\ref{Eq2} for $T_c(\Gamma_F)$ becomes:
\begin{equation}
T_c \approx T_{c0} - \frac{\pi\hbar}{4k_B} \frac{1}{2\tau_{tun,S}} \frac{\Gamma_F^2}{\Gamma_F^2+1},
\label{Eq7}
\end{equation}
in the thin-F regime $d_F \ll \ell_F$. We have reduced the parameter list to $R_b$, densities of states $2N_S(0)$ and $2N_F(0)$, and $E_{ex}$. Most notably, we have eliminated resistivities of F and S and the coherence length, $\xi_S$ in S. These are quantities whose measured in-plane values may not be the right values to use. 

From Eq.~\ref{Eq7}, the rapid decrease in $T_c$ at $\Gamma_F =1/\sqrt{3}$ crosses over to a slow decrease at $\Gamma_F \approx 1.5$. This crossover determines $E_{ex}$. Experimentally, the crossover occurs at a thickness $d_F$ less than the mean-free-path $\ell_F$. HM and Tagirov agree that in this regime motion back and forth through F is essentially ballistic, and the crossover occurs about when $2.6d_F \times$(number of hits needed to cross F/S interface)$=\xi_{F0}$. Anticipating that it takes several hits for a Cooper pair to get through the interface, we expect a theoretical fit to the data to find: $\xi_{F0} \approx 8d_F^{cr}$. Thus, just by seeing that the crossover in $T_c$ vs. $d_F$ occurs at $d_F \approx 1.5$ nm we immediately know that: $\xi_{F0} \approx 12$ nm. 

Most importantly, the crossover yields $E_{ex}$ in two steps. First, replacing $\Gamma_F\rightarrow 1.5$ and $d_F \rightarrow d_F^{cr}$ in Eq.~\ref{Eq6} yields: $2N_F(0)E_{ex}R_b\approx 0.75\hbar/d_F^{cr}e^2$. Second, given $\Gamma_F \approx 1.5$ at the crossover, the suppression of $T_c$ at the crossover is [Eq.~\ref{Eq8}]: $k_B(T_{c0} - T_c) \approx 0.7 \pi \hbar/16N_S(0)e^2d_F^{cr}R_b$, which yields $2N_S(0)R_b$. Obviously, the ratio of fit parameters, $2N_F(0)E_{ex}R_b/2N_S(0)R_b$ determines $E_{ex}$ to the extent that the ratio $2N_F(0)/2N_S(0)$ is known. Below we argue that a reasonable lower limit on this ratio is unity for Nb and $3d$ transition element ferromagnets.

We now turn to the final thick-F crossover where $T_c$ becomes completely independent of $d_F$. Physically, this crossover occurs at the thickness where nearly every Cooper pair that encounters the back of F is broken before it can get back to the F/S interface even once. In the clean-F regime of theory, this occurs about when $2\sqrt{2}d_F$ exceeds $\xi_{F0}$. In the dirty-F regime, $\ell_F \ll \xi_{F0}$, that we believe is more relevant to experiment, Cooper pairs that hit the backside of F get there by diffusion. In that case, the time required to get to the back of F and then back to the F/S interface is: $\Delta t \approx (2d_F)^2/D_F$. Thus, the dephasing angle, $\Gamma_F = (E_{ex}/\hbar)(2d_F)^2/D_F$ reaches $\pi$ just about when $d_F$ reaches $\xi_F$. Mathematically, this crossover lies in the $tanh$ in Eq. \ref{Eq3}, which approaches unity as $d_F$ surpasses $\xi_F$. 

The origin of the single shallow minimum that is sometimes seen in $T_c \text{~vs.~} d_F$ is interesting. Oscillations in $T_c$ due to spatial oscillations in the order parameter are damped by scattering, and they are not responsible. Because $\Gamma_F$ increases monotonically with $d_F$, it seems that $T_c$ should decrease monotonically. However, there is a change in boundary condition in the F layer that changes slightly the dependence of the pair-breaking rate on $\Gamma_F$. When F is thin, Cooper pairs return to the F/S interface by bouncing off of the backside of F; when F is thick, any unbroken Cooper pairs that return to the F/S interface get there by bouncing off of impurities. The new form is found from Eq.~\ref{Eq3} at $d_F \gg \xi_F$, where $\tanh(\sqrt{2i}d_F/\xi_F) \approx 1$:   
\begin{equation}
\Re\left\{\frac{1}{\tau_S}\right\} =\Re\left\{\frac{\Gamma_F}{(\Gamma_F+1-i)}\frac{1}{2\tau_{tun}}\right\}\\
=\frac{1}{2\tau_{tun,S}}\frac{\Gamma_F (\Gamma_F+1)}{(\Gamma_F+1)^2+1}.
\label{Eq8}
\end{equation}
For $\Gamma_F>1$, this expression yields a slightly smaller pairbreaking rate than Eq.~\ref{Eq7}, hence there is a slight increase in $T_c$ as $\Gamma_F$ increases to its ultimate, thick-F value. 

The most important part of the foregoing section is that theory provides a simple way to obtain $E_{ex}$ from the crossover thickness where the rapid drop in $T_c$ vs. $d_F$ finishes, regardless of whether one thinks of Cooper pairs as comprising time reversed electrons or equal energy electrons on the up and down electron Fermi surfaces. Moreover, the clean-F coherence length, $\xi_{F0}$, is a least several times longer than the thickness $d_F$ where the crossover occurs. 

\section{EXPERIMENTAL RESULTS AND DISCUSSION}

\begin{figure} [ht]
\begin{center}
\includegraphics[width=3.4in]{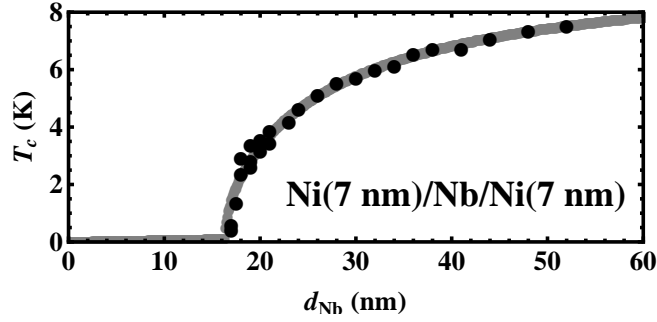}
\caption{$T_c$ vs. $d_{Nb}$ for Ni/Nb/Ni trilayers with thick Ni layers. \cite{Moraru2006} The good theoretical fit (solid curve) confirms that $1/2\tau_{tun,S} \propto 1/d_S$. The best fit finds a reasonable value for the specific resistance of the F/S interface: $R_b = 2.7\ \text{\emph{f}}\Omega\cdot \text{m}^2$. \label{Pairbreaking}}
\end{center}
\end{figure}

HM theory assumes that S is thinner than the superconducting coherence length, $\xi_S$. Rather than compare thickness with the measured $\xi_S$, which we believe is artificially shortened by surface scattering, etc., we rely on experimental evidence to argue that our films are thin enough that the theory applies. We do this by showing that the tunneling pair-breaking rate given in Eq.~\ref{Eq4} applies for Nb films as thick as ours. Figure \ref{Pairbreaking} shows $T_c \text{~vs.~} d_S$ for symmetric Ni/Nb/Ni trilayers with $17\ \text{nm} \le d_S \le 52\ \text{nm}$, from Moraru et al.\cite{Moraru2006} We fit the data by treating the trilayer as two independent bilayers, each with S thickness equal to half the Nb film thickness, and each in contact with an infinitely thick Ni film. We use the full theory, Eqs. \ref{Eq1} and \ref{Eq3}. The good fit (gray curve) confirms the qualitative prediction that $1/2\tau_{tun,S}\propto 1/d_S$, as expressed in Eq.~\ref{Eq4}, over the entire experimental range of thicknesses. 

Quantitatively, the best-fit value, $R_b = 2.7\ \text{\emph{f}}\Omega\cdot \text{m}^2$, compares well with the value, $R_b = 3.5\ \text{\emph{f}}\Omega\cdot \text{m}^2$, obtained by direct measurement of the Nb/Ni interface resistance.\cite{Fierz1990} We used $\Gamma_F = 4$ in the fit, per arguments in the preceding section, but the best-fit value of $R_b$ is insensitive to this choice as long as $\Gamma_F \gg 1$. The fitted curve in Fig. \ref{Pairbreaking} comes from Eq.~\ref{Eq1} with: $1/\tau_S = \frac{1}{2\tau_{tun,S}}\left(\frac{4}{5 - i}\right)$, and $T_{c0}=9.7$ K. We used a density of states in Nb: $2N(0) = 0.8 \times 10^{29}/\text{eV}\cdot \text{m}^3$.\cite{Mattheiss1970,DeVries1988,Jani1988} 
\FloatBarrier
\begin{figure} [h]
\begin{center}$
\begin{array}{c}
\includegraphics[width=3.4in]{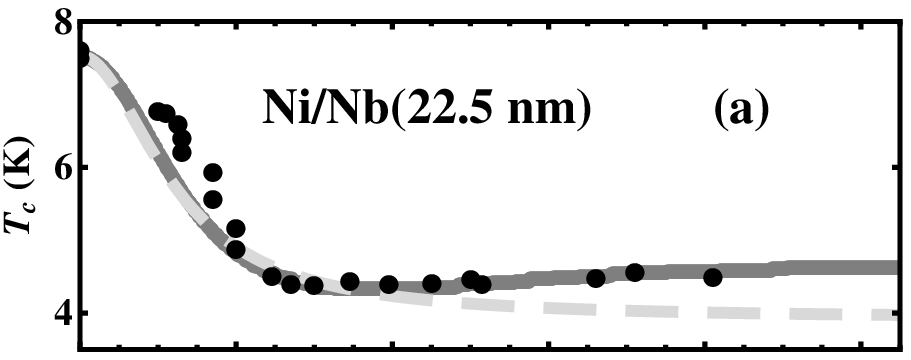} \\
\includegraphics[width=3.4in]{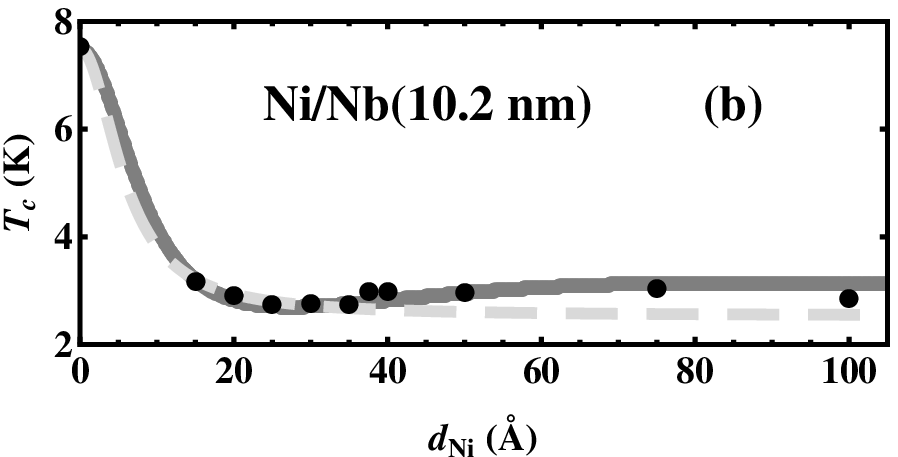}
\end{array}$
\end{center}
\caption{$T_c$ vs. $d_{Ni}$ for Ni/Nb bilayers found in (a) Kim et. al. \cite{Kim2005} and (b) Lemberger et. al \cite{Lemberger2008}. Values for exact [solid], (approximate) [dashed] fits are (a) $R_b = 2.1(3.0)\ \text{\emph{f}}\Omega\cdot \text{m}^2$, $2N_F(0)E_{ex} = 0.60 (0.56) \text{~nm}^{-3}$, $\xi_F=4.0$ nm; and (b) $R_b = 3.8(4.8)\ \text{\emph{f}}\Omega\cdot \text{m}^2$, $2N_F(0)E_{ex} = 0.60 (0.72) \text{~nm}^{-3}$, and $\xi_F=3.5$ nm.  \label{Ni-Nb}} 
\end{figure}

 F/Nb bilayers are deposited by dc sputtering from 2 in. diameter Nb, Ni, Py (Ni$_{0.8}$Fe$_{0.2}$), and Co$_{0.5}$Fe$_{0.5}$ targets onto oxidized Si substrates located 2 in. above the target. Si substrates approximately 18 x 18 x 0.4 mm$^3$ are placed into a load-locked UHV chamber with a base pressure of 5 x 10$^{-10}$ torr. In rapid succession, a buffer layer of Ge (10.5 nm) followed by a Nb layer and an F layer, and a protective layer of Ge (20 nm) are deposited. The deposition rates for Nb, Ge, CoFe, Py, and Ni are 1.5, 2.0, 1.30, 1.62, and 0.94 \AA/s, respectively. The Ge buffer layer on the oxidized silica improves reproducibility while the top Ge serves to inhibit oxidation in air. Substrates are nominally at room temperature during deposition. 
 
To obtain the $T_c$'s of our bilayers, we measure superfluid density as a function of temperature using mutual inductance of coaxial coils located on opposite sides of the sample at low frequency, 50 kHz. The coils are solenoids nominally 2 mm in diameter and 2 mm in length, much smaller than the 18 mm areal dimensions of the films. Various steps involved in converting mutual inductance to sheet conductivity have been described.\cite{Turneaure1996, Turneaure1998} With the resulting graphs values of $T_c$ are obtained using a quadratic fit near $T = T_c$, examples which can be found in previous publications.\cite{Lemberger2007, Lemberger2008}

Figure \ref{Ni-Nb} shows $T_c\text{~vs.~}d_F$ for Ni/Nb bilayers: (a) Ni/Nb(22.5 nm) \cite{Kim2005} and (b) Ni/Nb(10 nm). \cite{Lemberger2008} Note the shallow dip extending from $d_{Ni} \approx 2\ \text{nm}$ up to $d_{Ni} \approx 6$ nm in the former. To highlight the insensitivity of fit parameters to details, we analyze $T_c$ vs. $d_F$ in two ways. One uses the highly-simplified two-parameter, thin-F, small-pair-breaking version of theory embodied in Eqs. \ref{Eq4}, \ref{Eq6}, and \ref{Eq7} (dashed curve in this figure, and in subsequent figures). The other (solid curves in the figures) uses the full theory (Eqs. \ref{Eq1} and \ref{Eq3}) and includes the third fit parameter, $\xi_F$. Fit parameters are given in the figure captions and in Table \ref{Table1}. Best-fit values of $R_b$ for Nb/Ni are $2.1$ and $3.8\ \text{\emph{f}}\Omega\cdot \text{m}^2$, for data from Kim et al.\cite{Kim2005} and Lemberger et al.,\cite{Lemberger2008} respectively, both consistent with the fit in Fig. 1. Given $R_b$, fitted values of $2N_{Ni}(0)E_{ex}$ are $0.60$ and $0.88/\text{nm}^3$.

\begin{figure} [ht]
\begin{center}
\includegraphics[width=3.4in]{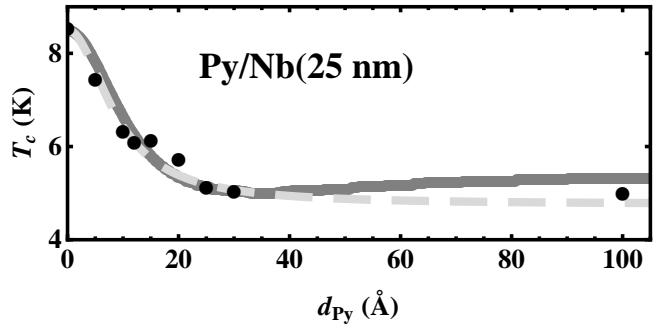}
\caption{$T_c$ vs. $d_{Py}$ for our Py/Nb bilayers. Fitting values for exact [solid], (approximate) [dashed] fits are  $R_b = 1.9(2.6)\ \text{\emph{f}}\Omega\cdot \text{m}^2$, $2N_F(0)E_{ex} = 0.88 (0.88) \text{~nm}^{-3}$, and $\xi_F=4.0$ nm. The lack of data in the flat regime gives us less, though still reasonable, confidence in $\xi_F$.  \label{Py-Nb}}
\end{center}
\end{figure}

\begin{figure} [h]
\begin{center}$
\begin{array}{c}
\includegraphics[width=3.4in]{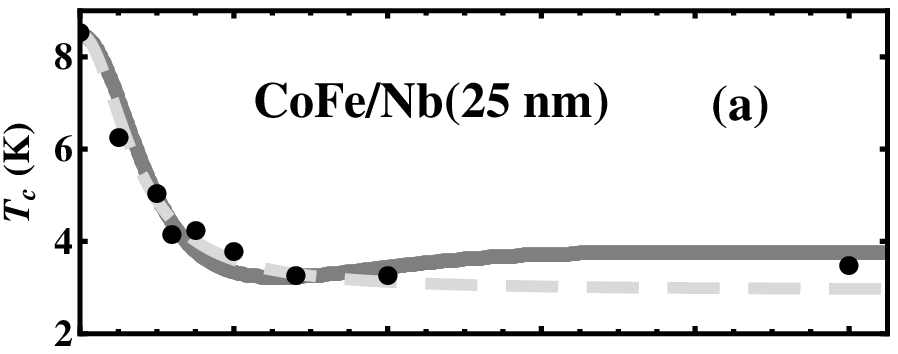} \\
\includegraphics[width=3.4in]{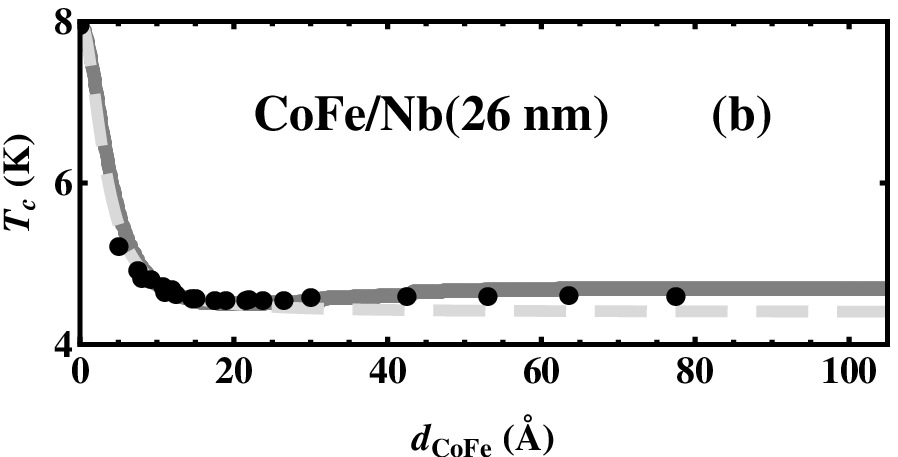}
\end{array}$
\end{center}
\caption{$T_c$ vs. $d_F$ for CoFe/Nb from (a) the present work, and (b) Kim et al. Fitting values for exact [solid], (approximate) [dashed] fits are (a) $R_b = 1.3(1.7)\ \text{\emph{f}}\Omega\cdot \text{m}^2$, $2N_F(0)E_{ex} = 1.5 (1.6) \text{~nm}^{-3}$, $\xi_F=3.0$ nm; and (b) $R_b = 2.1(2.6)\ \text{\emph{f}}\Omega\cdot \text{m}^2$, $2N_F(0)E_{ex} = 2.3 (2.4) \text{~nm}^{-3}$, $\xi_F=3.0$ nm.\label{CoFe-Nb}}
\end{figure}

Bilayers involving strongly ferromagnetic $3d$ alloys Py and CoFe yield similar results, Figs. \ref{Py-Nb} and \ref{CoFe-Nb}a. For Py/Nb, best-fit values are: $R_b = 1.9\ \text{\emph{f}}\Omega\cdot\text{m}^2$, $2N_{Py}(0)E_{ex} = 0.88/\text{nm}^3$, and $\xi_F=4.0$\ nm. $\xi_F$ is somewhat more uncertain due to lack of data above the minimum, though it seems reasonably apparent.

For CoFe/Nb bilayers, Fig. \ref{CoFe-Nb}, the shallow minimum in $T_c$ vs. $d_{CoFe}$ seems to be present in our data, Fig. \ref{CoFe-Nb}a, and in the data of Kim et al., Fig. \ref{CoFe-Nb}b. The initial drops in $T_c$ and the ultimate values of $T_c$ are different in the two data sets for unknown reasons. The detailed fits to our data and those of Kim et al. find $R_b = 1.3$ and $2.1\ \text{\emph{f}}\Omega\cdot\text{m}^2$, respectively, about half the value for Ni/Nb bilayers. Values of $2N_{CoFe}(0)E_{ex}$ differ by 50\%, while values of $\xi_F$ are the same, Table \ref{Table1}.  

Now we turn to the key quantitative issue - the effective exchange energy. To get experimental values for $E_{ex}$, we need to interpret the total density of states, ``$2N_F(0)$'', that appears in the free-electron theory. Two choices come to mind. In a theory that included ferromagnetic conduction bands, we might expect to find that $2N_F(0)$ is replaced by the total density of states, $N_{F\uparrow} (0) + N_{F\downarrow} (0)$, of majority and minority spin densities of states. Experimental total densities of states for Fe, Co, and Ni range from 1.54 to 2.07 / eV-atom. \cite{Coey2010} Thus, at an accuracy appropriate for present purposes, the total density of states is about $1.8\pm0.3$/eV$\cdot$atom $= 1.7\pm0.3 \times 10^{29}/$eV$\cdot$m$^3$ for all three F layers. On the other hand, it is conceivable that in a better theory the smaller of $N_{F\uparrow} (0)$ and $N_{F\downarrow} (0)$ would create a bottleneck of sorts, and ``$2N_F(0)$'' would be replaced by two times the smaller of $N_{F\uparrow} (0)$ and $N_{F\downarrow} (0)$. Tunneling \cite{Tedrow1994} and point-contact Andreev reflection\cite{Soulen1998, Upadhyay1998} experiments both find that the ratio of larger to smaller density of states in $3d$ ferromagnetic metals is about 2.4$\pm$0.3. In this case, ``$2N_F(0)$'' would be about half of the total density of states, i.e., about $0.8\pm0.15 \times 10^{29}/$eV$\cdot$m$^3$. We take this as a reasonable lower limit on the effective total density of states in F, which yields an upper limit on $E_{ex}$. Authors that have found exchange energies much larger than those reported here have, in effect, used an F density of states much smaller than that used here, among other differences.

\begin{table}
\caption{Parameters obtained from theoretical fits to $T_c$ vs. $d_F$ for three F/S bilayer systems. Fermi velocities $v_{FF}$ in F are from the literature. [See text for details.] $R_b$ was obtained from fit parameter $2N_S(0)R_b$ by using $2N_S(0) = 0.8\times10^{29} /\text{eV}\cdot \text{m}^3$ for Nb. \cite{Mattheiss1970,DeVries1988,Jani1988} Values of $R_b$ in parentheses are taken from the simple, approximate fits in the figures. The effective exchange energy, $E_{ex}$, was obtained from $2N_F(0)E_{ex}$ using $2N_F(0) = 0.8 \times 10^{29}/\text{eV}\cdot \text{m}^3$. [See text.]}

\label{Table1}
\begin{center}
 \begin{tabular}{| c | c | c | c | c | c | c | c | }
  \hline
 F/S & Ref. & $R_b$ & $2N_F(0)E_{ex}$ & $\xi_F$ & $v_{FF}$ & $E_{ex}$& $\xi_{F0} = \frac{\hbar v_{FF}}{E_{ex}}$ \\
 & & $\left[\emph{f}\Omega\cdot\text{m}^2\right]$ & $\left[1/\text{nm}^3\right]$ & $\left[\text{nm}\right]$ & $\left[10^6\text{ m/s}\right]$ & $\left[\text{K}\right]$ $\left[\text{meV}\right]$ & $\left[ \text{nm}\right]$ \\ \hline
Ni/Nb(10.2 nm) & TRL & 3.75 (4.8) & 0.60 (0.72) & 3.5 & 0.28 & 87~~~~7.5 (9) & 25\\ \hline 
Ni/Nb(22.5 nm) & Kim & 2.05 (3.0) & 0.60 (0.56) & 4.0 & 0.28 & 87~~~~7.5 (7) & 25\\ \hline
Py/Nb(25 nm) & Hinton & 1.85 (2.6) & 0.88 (0.88) & 4.0 & 0.22 & 128~~~~11 (11) & 13\\ \hline
CoFe/Nb(25 nm) & Hinton & 1.25 (1.7) & 1.5 (1.6) & 3.0 & 0.33 & 220~~~~19 (20) & 11\\ \hline 
CoFe/Nb(26 nm) & Kim & 2.1 (2.6) & 2.3 (2.4) & 3.0 & 0.33 & 337~~~~29 (30) & 7.5\\ \hline
 \end{tabular}
 \end{center}
 \end{table}

Table \ref{Table1} shows that effective exchange energies are about a factor of five smaller than $k_BT_C$, [$T_C = 627$ K, \cite{Ashcroft1976} 871 K, \cite{Wakelin1953} and 1600 K, \cite{Lezaic2007} for Ni, Py, and CoFe, respectively.] In fact, the values for TRL/Hinton films are each approximately 14\% the quoted $T_C$. This is our main result. For comparison, exchange energies obtained from ARPES are several times larger than $k_BT_C$, about 0.25 eV (2900 K) for Ni and Py, and more than 0.5 eV ($> 5800$ K) for Co.\cite{Petrovykh1998} 

One naturally wonders whether our small experimental exchange energies imply an unreasonable value of some other quantity. Consider the transparency of the F/S interface. If we use Fermi velocities from ARPES (given in the Table; experimental values are the same for majority and minority spins), we find: $\xi_{F0} \approx 25$ nm for Ni, 13 nm for Py, and 11 nm for CoFe. Given that the first crossover in $T_c$ vs. $d_F$ occurs at $d_F^{cr} \approx 1.5$ nm for all three bilayers, these values of $\xi_{F0}$ imply that Cooper pairs must hit $N \approx \xi_{F0}/3d_F^{cr}$ times, i.e.: 6, 3, and 2 times for Ni/Nb, Py/Nb, and CoFe/Nb interfaces, respectively, in order to get through. Alternatively, we can estimate N from: $N \approx 3R_b/2\rho_F \ell_F$, a relationship discussed above, with $\rho_F \ell_F=3/2N_F(0)v_{FF}e^2$. We find: $N =$ 7, 3, and 3, respectively, for Ni/Nb, Py/Nb, and CoFe/Nb. The factor-of-two difference in Fermi velocities between S and F guarantees that more than one hit is necessary, so these numbers seem consistent with realistic, clean interfaces. 

In this regard, we note that Aarts and Geers studied Fe/Nb/Fe trilayers \cite{Geers2001} and V/V$_{1-x}$Fe$_x$ multilayers. \cite{Aarts1997} They also concluded that transmission probabilities were significantly smaller than unity. Aarts et al.,\cite{Aarts1997} found a high transparency for the V/VFe interface, but only when V$_{1-x}$Fe$_x$ was mostly V.

Consider the electron mean free path $\ell_F$ in F. Sidorenko et al. obtain $\ell_F \approx$ 1.8 nm in Ni. The short mean-free-path is necessary to account for the small amplitude of the oscillation in $T_c$ that they observe. They offer an explanation for why the effective $\ell_F$ deduced from $T_c$ vs. $d_F$ might be much shorter than values commonly cited for $3d$ metals, which they note range from 5 nm to 30 nm.\cite{Gurney1993,Dieny1994} We also find short mean-free-paths, obtained from our experimental $\xi_F$: $\ell_F \approx 3\xi_F^2/\xi_{F0} =$ 1.5 , 3.7, and 2.5 nm for Ni, Py and CoFe, respectively. In our case, the short $\ell_F$ is necessary to account for the minimum in $T_c$ occurring at such a small thickness, $d_F$. Our use of HM theory relies on F being ``dirty'': $\ell_F < \xi_{F0}$, which is satisfied.

Sidorenko et al. deduce $\xi_{F0} =$ 0.88 nm in Ni from their fit. They do not give an exchange energy, but if we use $v_F = 0.28 \times 10^6$ m/s for Ni, this implies $E_{ex} = 200$ meV, which is much larger than we find (16 meV) when we fit their data. Part of the reason for the difference is that they effectively used a density of states for Ni that is about six times smaller than the density of states in Nb, and the crossover in their fit occurs at $d_F \approx 0.3$ nm, while our fit hits the experimental crossover at $d_F =$ 1.2 nm. Kim et al. obtain exchange energies four times higher than we get when we fit their data because they effectively used a density of states for F that is about four times smaller than their density of states for Nb. 

\section{CONCLUSIONS}

Using the standard free-electron theory of the F/S proximity effect to interpret $T_c$ vs. $d_F$, plus literature values for densities of states and Fermi velocities for Nb and $3d$ ferromagnets, we conclude that effective exchange energies for dephasing of Cooper pairs in strong $3d$ ferromagnets is 15 times smaller than experimental exchange energies from ARPES, and five times smaller than the value deduced from previous measurements of $T_c$ vs. $d_F$. It may be coincidental, but exchange energies for Ni, Py, and CoFe scale with Curie temperatures. The clean-limit coherence length for Cooper pairs in F is about ten nanometers, much larger than previously thought. While it seems that the thickness, $d_F^{cr}$, at the crossover where the rapid decrease of $T_c$ gives way to flat behavior should be the characteristic length scale for Cooper pairs in F, in fact the characteristic length scale, $\xi_{F0}$, is the typical distance traveled by a Cooper pair in F before returning to S, which is the back-and-forth transit distance ($\approx 3d_F^{cr}$) for a pair multiplied by the times it must bump into the F/S interface before getting back into S. Thus, $\xi_{F0}$ is typically a factor of 5 to 10 larger than $d_F^{cr}$. 

That $E_{ex}$ is smaller than previously thought has implications for interpretation of other measurements, e.g. Josephson coupling vs. $d_F$ in S/F/S' trilayers. 

A possible reason for weakened pair-breaking is spin-orbit scattering, which has been considered theoretically. \cite{Demler1997,Buzdin2005} The effect of the exchange energy on Cooper pairs is essentially the same as the pair-breaking Zeeman effect of a magnetic field, $B$, on Cooper pairs, which has a pair-breaking rate, $1/\tau_S = 2\mu_B B/\hbar$, when spin-orbit scattering is negligible. But when spin-orbit scattering is strong, the pair-breaking rate is reduced by a factor of $\mu_B B/(\hbar / \tau_{so})$. \cite{Tinkham1975} We propose that this mechanism is part of the explanation, with $E_{ex}$ in place of $2 \mu_B B$. 

Finally, our analysis finds that Cooper pairs must strike the F/S interface several times before getting through. The CoFe/Nb and Py/Nb interfaces are a little more transparent than the Ni/Nb interface.  

\begin{acknowledgments}
This work was supported in part by NSF grant DMR-0805227. We acknowledge useful discussions with Jan Aarts, Norman Birge, Julia Meyer, and Christoph Strunk.
\end{acknowledgments}

\FloatBarrier

\bibliography{bibliography}

\end{document}